\documentclass[iop]{emulateapj}

\def\lapp{\ifmmode\stackrel{<}{_{\sim}}\else$\stackrel{<}{_{\sim}}$\fi}
\def\gapp{\ifmmode\stackrel{>}{_{\sim}}\else$\stackrel{>}{_{\sim}}$\fi}
\usepackage{multirow}
\usepackage{color}
\usepackage{amsmath}
\usepackage{soul}
\usepackage{mathrsfs}
\begin{document}

\title{Light curve and SED modeling of the gamma-ray binary 1FGL~J1018.6$-$5856:
constraints on the orbital geometry and relativistic flow}

\author{
Hongjun An\altaffilmark{1,2} and Roger W. Romani\altaffilmark{1}
\\
{\small $^1$Department of Physics/KIPAC, Stanford University, Stanford, CA 94305-4060, USA}\\
{\small $^2$Department of Astronomy and Space Science, Chungbuk National University, Cheongju, 28644, Republic of Korea}\\
}
\altaffiliation{$^3$hjan@chungbuk.ac.kr}

\begin{abstract}
	We present broadband spectral energy distributions (SEDs) and light curves
of the gamma-ray binary 1FGL~J1018.6$-$5856 measured in the X-ray and the gamma-ray bands.
We find that the orbital modulation in the low-energy gamma-ray band is similar to that in 
the X-ray band, suggesting a common spectral component. However, above a GeV the
orbital light curve changes significantly. We suggest that the GeV band contains significant
flux from a pulsar magnetosphere, while the X-ray to TeV light curves are 
dominated by synchrotron and Compton emission from an intrabinary shock (IBS).
We find that a simple one-zone model is inadequate to explain the IBS emission, but
that beamed Synchrotron-self Compton radiation from adiabatically accelerated
plasma in the shocked pulsar wind can reproduce the complex multiband light curves,
including the variable X-ray spike coincident with the gamma-ray maximum.
The model requires inclination $\sim$50$^\circ$ and orbital eccentricity $\sim$0.35,
consistent with the limited constraints from existing optical observations.
This picture motivates searches for pulsations from the energetic young pulsar
powering the wind shock.
\end{abstract}


\section{Introduction}

        High-energy gamma-ray emission in the GeV to TeV band has been observed
in several high-mass X-ray binaries. These so-called gamma-ray binaries are bright 
in all electromagnetic wavebands. Aside from the star-dominated optical, they show
significant orbital flux modulation across the electromagnetic spectrum.
High-energy emission of gamma-ray binaries is theorized
to be from the intrabinary shock (IBS, pulsar model) or the jet
\citep[black hole model, see][for reviews]{m12,d15}.
There are only a handful of these objects known, and in only one of them
is the compact source type securely identified \citep[PSR~B1259$-$63;][]{jml+92}.
However models of orbital flux modulation are important for constraining the
nature of the other sources. 

        Gamma-ray binaries PSR~B1259$-$63 and LS~I~+61$^\circ$~303 have Be star companions,
and their emission is modeled via episodic crossing of the compact object (pulsar) of
the equatorial outflow of the stellar companion \citep[][]{tak94}.
The other sources have O-star companions and are
explained via a wind-wind interaction or a microquasar model \citep[][]{bp04}.
The prototypical object in this class is LS~5039.
It has been extensively studied over all electromagnetic wavelengths, and its orbital
parameters are relatively well measured.

        1FGL~J1018.6$-$5856 (3FGL~J1018.9$-$5856, hereafter J1018) is another gamma-ray binary
with an O-star companion.
The source has similar properties to those of LS~5039 but is less
well studied because of its longer 16.5\,d orbital period and X-ray faintness.
Recently, modulated TeV emission was detected from the source \citep[][]{J1018hess15}.
Furthermore, X-ray and optical observations of the source were able to constrain
the nature of the compact object and the orbital parameters \citep[][]{wr15,abb+15,scc15,wrk+15},
making this a likely neutron star in a mildly eccentric binary.

        J1018 has several properties which challenge current spectral energy distribution (SED) emission models.
\citet{scc15} found that the orbital phase of the maximum gamma-ray flux
coincides with inferior conjunction (compact object is in front of the stellar
companion). This is puzzling because in these sources gamma rays
are believed to be produced via inverse-Compton up-scattering of the stellar
UV photons; the gamma-ray flux is expected to be maximum when the compact object is behind
because the collision geometry is favorable. In addition, the X-ray light curve of J1018 exhibits
two peaks, one being narrow and highly variable, and the other being broad and stable
in time \citep[][]{adkh13,abb+15}. The double-peaked X-ray light curve cannot be easily explained
with simple orbital modulation of the binary. Further studies of these intriguing properties
of J1018 can give us new insights into gamma-ray binaries.

        In this paper, we use archival IR/UV/X-ray data and a new {\it Fermi}
Large Area Telescope (LAT) analysis of J1018
to find a scenario that explains the observed properties. In Section~\ref{sec:sec2},
we describe the observations and the data reduction. We then present results of the data analysis and
modeling in Sections~\ref{sec:sec3} and \ref{sec:sec4}. The model-inferred orbital/physical parameters
are then compared with the observed and theoretical values to verify the model.
Finally, we discuss and conclude in Section~\ref{sec:sec5}.

\section{Observations and Data Reduction}
\label{sec:sec2}
	We use 7-yr {\it Fermi}-LAT data obtained between 2008 August 4 and 2015 August 27
to measure the gamma-ray properties of J1018.
The Pass 8 \citep[][]{fermiP8} processed data were downloaded from the {\it Fermi} Science
Support Center (FSSC),\footnote{http://fermi.gsfc.nasa.gov/ssc/data/analysis/documentation/P\\ass8\_usage.html}
then reduced and analyzed with the {\it Fermi}-LAT
Science Tools {\tt v10r0p5} along with the instrument response functions (irfs) P8R2\_V6.
We selected source class events with Front/Back event type in the 100\,MeV--500\,GeV band
using an $R=5^\circ$ circular region of interest (ROI),
and applied $<90^\circ$ zenith angle and $<52^\circ$ rocking angle cuts.

	For the UV band, we use archival {\it Swift}/UVOT \citep[][]{pbp+08} data taken between
MJD~55103 and MJD~56992. The source flux was calculated in the six {\it Swift}/UVOT bands
with the {\tt uvotsource} tool integrated in Heasoft~6.16 along with the HEASARC remote
CALDB\footnote{http://heasarc.nasa.gov/docs/heasarc/caldb/caldb\_remote\_ac\\cess.html}.
We used a 5$''$ and a 15$''$ aperture for the source and the background, respectively.
For other wavebands, especially the X-ray band,
we use catalog data\footnote{http://irsa.ipac.caltech.edu/frontpage/}
and previously reported results
\citep{J1018fermi12,adkh13,abb+15,J1018hess15}.

\section{Data Analysis and Results}
\label{sec:sec3}

\subsection{Orbital Modulation in the Fermi-LAT band}
\label{sec:sec3_1}
	With the new Pass 8 {\it Fermi}-LAT data, we verify the orbital period measured in
the X-ray and the gamma-ray bands \citep[][]{abb+15,cccc+14} using the epoch folding
method developed by \citet{l87}. Because the LAT's point spread function (PSF)
is broad and the source is in a crowded region, a large number of background events is expected.
We therefore weight each event with the probability that the event is from the source
using the {\tt gtsrcprob} tool (considering all the sources within $R=15^\circ$; see
Section~\ref{sec:sec3_2}).
We then selected events in a small aperture ($R \leq 0.5^\circ-2^\circ$)
to have good signal to background
ratio, and folded the probability-weighted event time series on various test periods
to produce orbital light curves.
In doing so, the exposure is separately calculated and folded on the same test periods,
and the light curves are corrected for exposure variations.
We calculated $\chi^2$ for a constant function for each
test period and fit the measured $\chi^2$'s
to find the best orbital period ($P_{\rm orb}$). We do this for various energy bands,
apertures and source probability thresholds (obtained with {\tt gtsrcprob}),
and find that the resulting $P_{\rm orb}$ is 16.539--16.555\,days which is
consistent with the X-ray measurement \citep[$P_{\rm orb}=16.544\pm0.008$\,days;][]{abb+15}.
Therefore, we use
$P_{\rm orb}=16.544$\,days and $T_{\rm 0}=55403.4$ MJD for $\phi_{\rm orb}=0$
(corresponding to the gamma-ray maximum and inferior conjunction) throughout this paper.

\begin{figure*}
\centering
\hspace{-0.2 in}
\includegraphics[width=7.1 in]{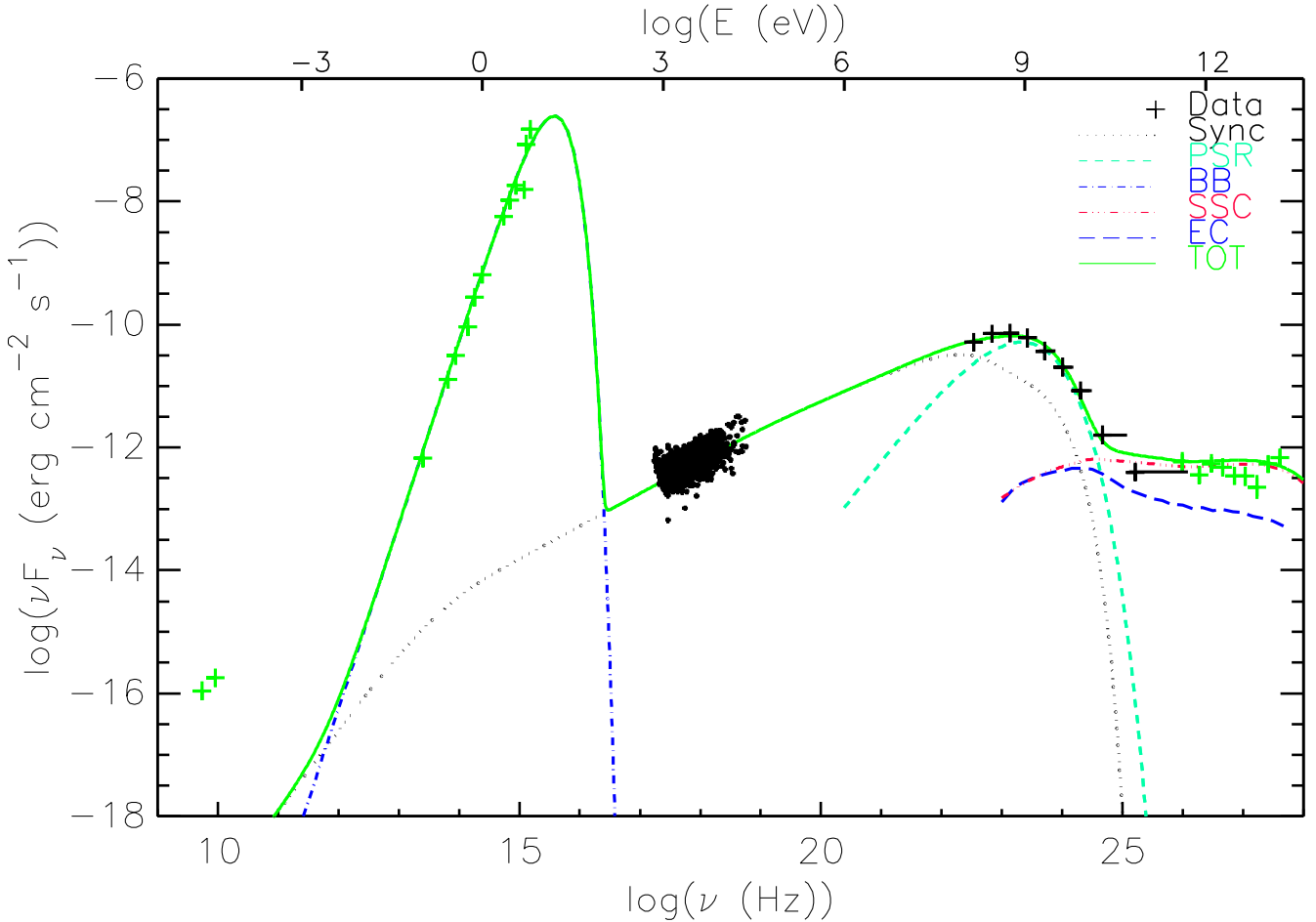}
\figcaption{Phase-averaged broadband SED and an emission model (Section~\ref{sec:sec4_2}).
Note that X-ray flux varies significantly over the phases, and we show all the X-ray data here.
All the {\it Fermi}-LAT data points are detected significantly with the TS values greater than 15
and the {\it H.E.S.S.} points are denoted in green for better visibility.
The model is calculated for 40 phase bins and is averaged.
The summed model is shown in solid green, and each component is also shown.
Note that we do not attempt to model the low-energy (presumably cooled, large-scale) electron
population producing the radio emission.
See Section~\ref{sec:sec4_2} and Table~\ref{ta:ta2} for the model and the parameters.
\vspace{0.0 mm}
\label{fig:fig2}
}
\end{figure*}

\subsection{Fermi-LAT data analysis}
\label{sec:sec3_2}
        We first measure the phase-averaged gamma-ray spectrum of the source using the
binned likelihood analysis with the {\it Fermi}-LAT {\tt gtlike} tool.
We used a 5$^\circ$ aperture and fit all the bright 3FGL sources \citep[][]{fermi3fgl}
within the aperture
(detected with confidence greater than 5$\sigma$) and the diffuse/isotropic emission
\citep[{\tt gll\_iem\_v06};][{\tt iso\_P8R2\_SOURCE\_V6\_v06}]{fermigllv06}.\footnote{
http://fermi.gsfc.nasa.gov/ssc/data/access/lat/BackgroundM\\odels.html}
We consider energy dispersion in the analysis.\footnote{http://fermi.gsfc.nasa.gov/ssc/data/analysis/scitools/binned\_l\\ikelihood\_tutorial.html}
J1018's emission is modeled with the log-parabola function
$dN/dE=N_{\rm 0}(E/E_{\rm b})^{-\alpha - \beta\mathrm{log}(E/E_{\rm b})}$.
After this study, we gradually freeze parameters for faint sources until the
fit statistic \citep[$k - \mathrm{log} \mathscr{L}$;][]{aic74} is minimized in order to remove unnecessary fit parameters.
We find that the central values for the best-fit parameters for J1018 do not change in this case, and
the best fit is obtained when we fit parameters for five bright sources and the diffuse/isotropic emission.
J1018 is detected with high significance (test statistic $TS=14520$) and
the new best-fit parameters are very similar to those reported in the
3FGL catalog \citep[][]{fermi3fgl}. We verified that the results (Table~\ref{ta:ta1}) agree
with those measured using a large aperture (e.g., $R=15^{\circ}$).

We then estimated systematic uncertainties due to variations of effective area and the interstellar
diffuse model (gll\_iem\_v06). We varied the effective area using the bracketing
scales\footnote{http://fermi.gsfc.nasa.gov/ssc/data/analysis/scitools/Aeff\_Sys\\tematics.html\#bracketing}
and the normalization of the interstellar diffuse emission by $\pm$6\%, and performed {\tt gtlike} analysis as we
did above. The measured SED is shown in Figure~\ref{fig:fig2},
and the results of the likelihood fit are presented in Table~\ref{ta:ta1}.

\newcommand{\markaa}{\tablenotemark{a}}
\newcommand{\markbb}{\tablenotemark{b}}
\begin{table}[t]
\vspace{-0.0in}
\begin{center}
\caption{Fit results for the 100\,MeV--500\,GeV phase-averaged {\it Fermi}-LAT data with
a log-parabola model
\label{ta:ta1}}
\vspace{-0.05in}
\scriptsize{
\begin{tabular}{ccc} \hline\hline
parameter			& units	& value\markbb  \\ \hline
$\alpha$			& $\cdots$	&  $2.496\pm0.013\pm0.015$  \\
$\beta$				& $\cdots$	&  $0.229\pm0.009\pm0.011$  \\
$E_{\rm b}$\markaa		& GeV		&  1350.210  \\
$F_{\rm 100\,MeV-500\,GeV}$	& $\rm ph\ cm^{-2}\ s^{-1}$	& $3.67\pm0.08\pm0.12\times 10^{-7}$ \\ \hline
\end{tabular}}
\end{center}
\vspace{-0.5 mm}
$^{\rm a}${Fixed.}\\
$^{\rm b}${Statistical and systematic uncertainties are reported.
For the systematic uncertainties, those of the LAT response functions
and of the interstellar emission model are summed in quadrature.}\\
\end{table}
 
\begin{figure*}
\centering
\begin{tabular}{ccc}
\hspace{-16.0 mm}
\includegraphics[width=4.25 in]{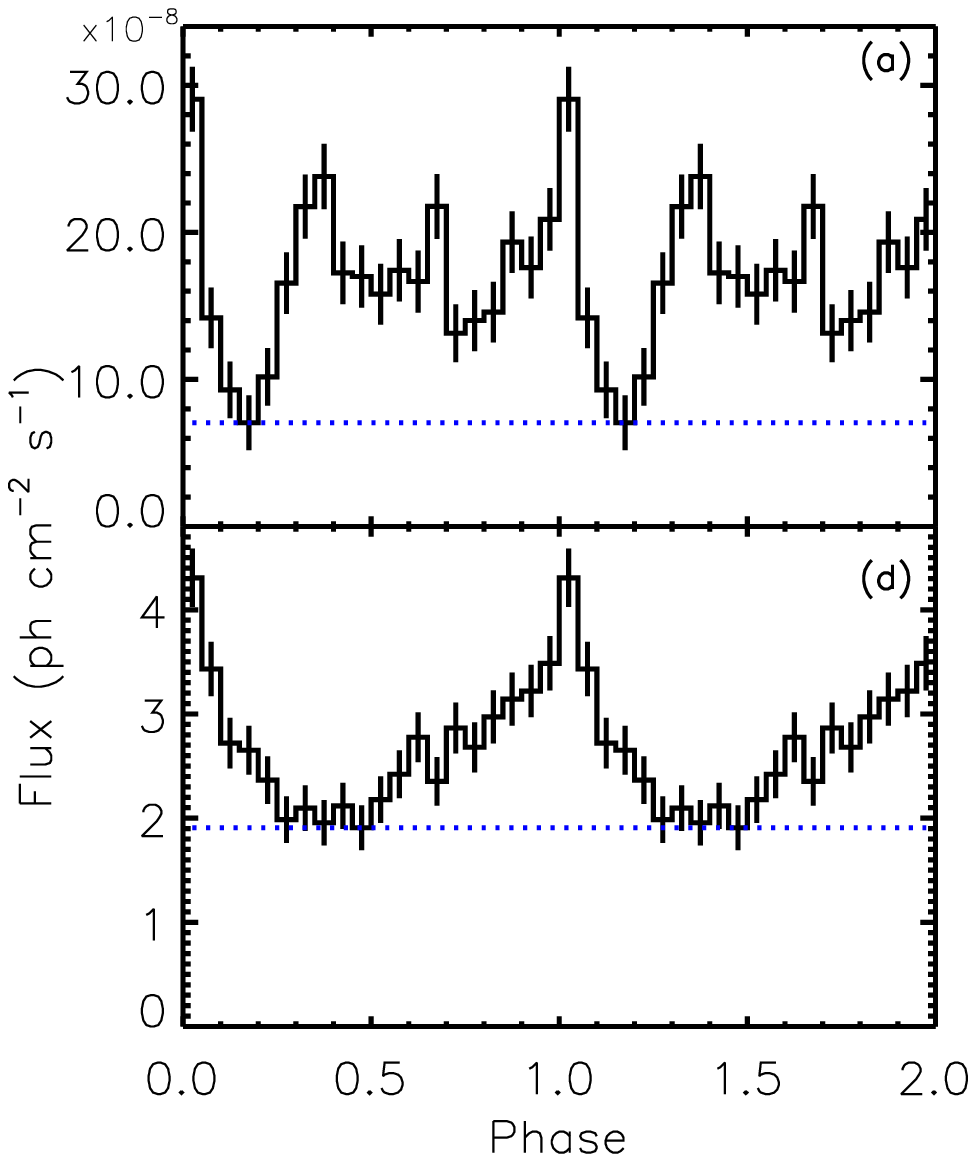} &
\hspace{-52.0 mm}
\includegraphics[width=4.25 in]{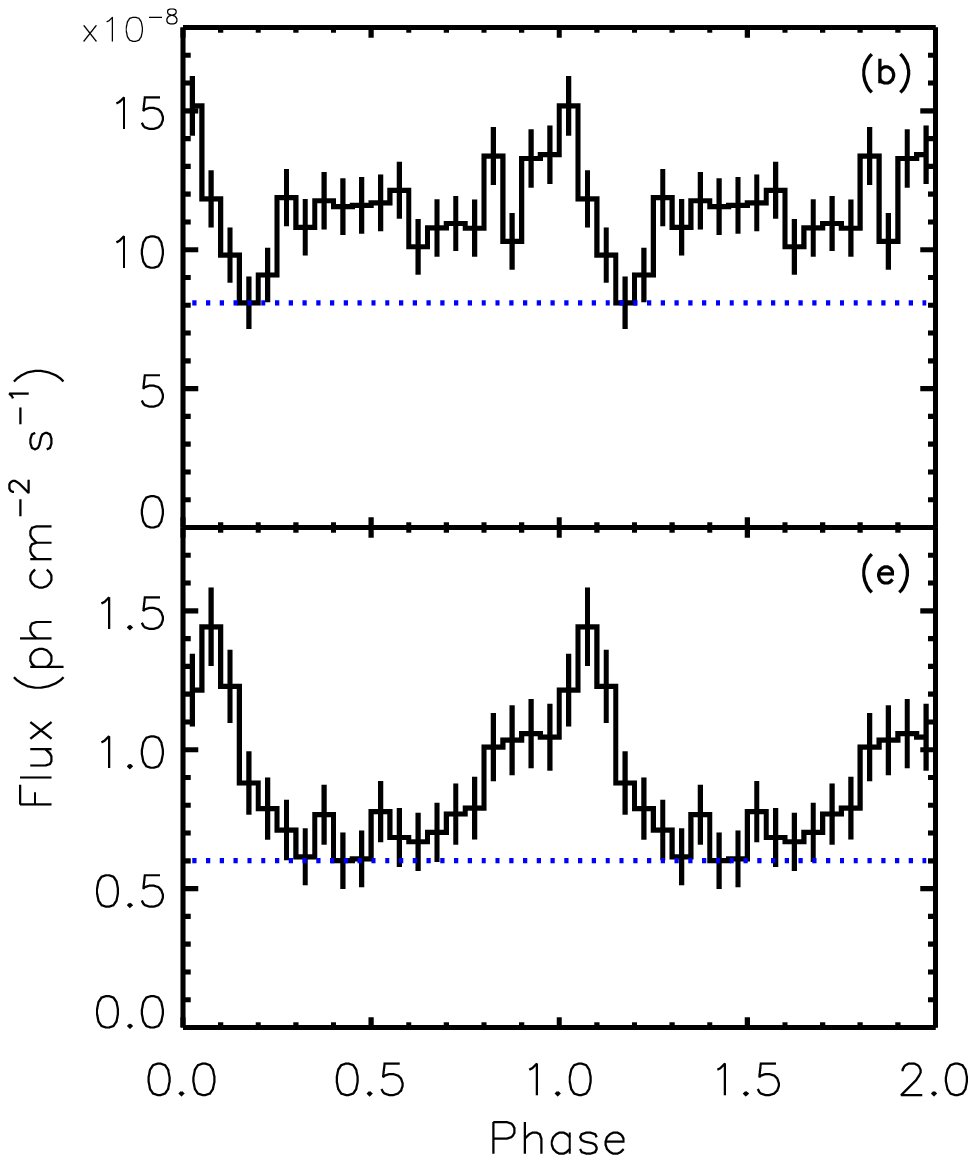} &
\hspace{-52.0 mm}
\includegraphics[width=4.25 in]{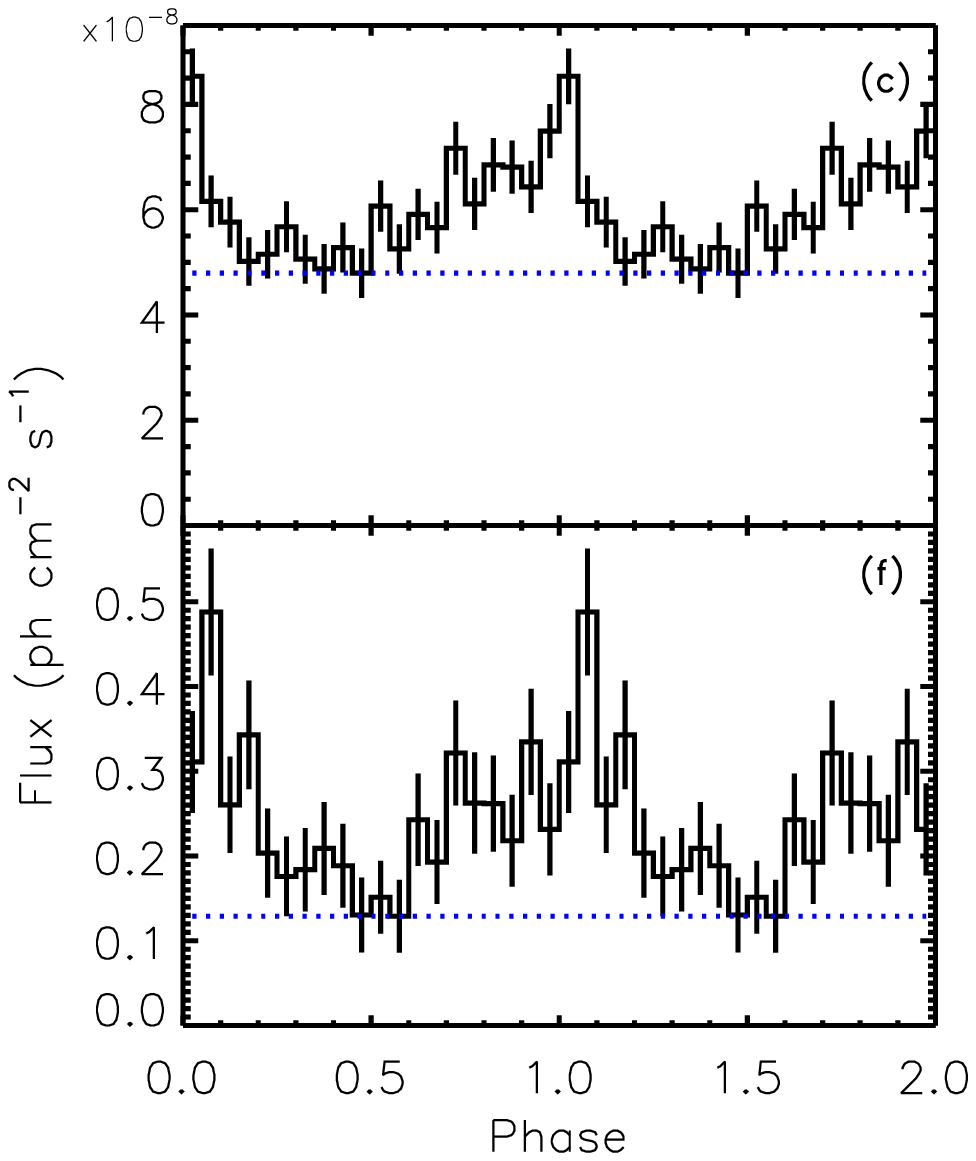} \\
\end{tabular}
\vspace{-5.0 mm}
\figcaption{Energy-resolved {\it Fermi}-LAT light curves. Flux is shown as a function of orbital
phase. The energy bands are (a) 100--200\,MeV, (b) 200--400\,MeV,
(c) 400-800\,MeV, (d) 800--1600\,MeV, (e)1600--3200\,MeV,
and (f) 3200--6400\,MeV. The blue line shows the minimum flux level.
Note that variability is minimal at 
intermediate gamma-ray energies. No significant modulation is
seen above 6400\,MeV due to the paucity of counts.
\label{fig:fig3}
}
\vspace{0mm}
\end{figure*}

              We then performed a phase-resolved spectral analysis on 20 orbital
phase bins to measure spectral variability of the source. We folded the data on the orbital
period ($P_{\rm orb} = 16.544$\,days) and selected events for each of the 20 phase bins.
We performed a binned likelihood analysis using the same parameters for the phase-averaged
analysis above, but held all background (diffuse and source) parameters fixed at the 
phase-averaged values, varying only those for J1018.
For each of the 20 phase bins, we generated an SED. These were used to produce energy-resolved
orbital light curves (Figure~\ref{fig:fig3}).
The low-energy $<$400\,MeV light curves show structure similar to
that seen in the X-ray band; there is a sharp peak at phase 0, and a relatively broad
hump at phases 0.2--0.7 (with possibly peaked substructure). 
At higher LAT energies the broad hump disappears, resulting in a
light curve resembling that measured at very high energy (VHE) with
{\it H.E.S.S.} \citep[][]{J1018hess15}.
We infer that a separate component contributes to the {\it Fermi}-LAT band below  $\sim$400\,MeV.

\subsection{Constructing broadband SEDs}
\label{sec:sec3_3}
	We assemble the spectral energy distribution of the source in the radio to the TeV
band. The radio, X-ray and TeV data are taken from \citet{J1018fermi12},
\citet{abb+15}, and \citet{J1018hess15}, respectively. We also use the IR-band flux taken
from the {\it WISE} and the {\it 2MASS} catalogs.
For the {\it Swift}/UVOT data we verified that there is no orbital flux
modulation. Therefore, we take the phase-averaged UV flux for the SED.
The IR-to-UV magnitudes are properly converted into fluxes
and corrected for extinction \citep[$E(B-V)=1.35$;][]{wr15} with the \citet{sf11} calibration.
Note that the IR-to-UV data are well described by a blackbody model having $kT=3.8$\,eV and
bolometric luminosity $9\times10^{38}\rm \ erg\ s^{-1}$, typical for an O star
(Figure~\ref{fig:fig2}). The phase-averaged {\it Fermi}-LAT SED was produced as described above
(Section~\ref{sec:sec3_2}), and we show the broadband SED in Figure~\ref{fig:fig2}.
Note that the X-ray spectrum varies significantly depending on the orbital phase \citep[][]{adkh13}.

         In Figure~\ref{fig:fig2}, we also show an emission model (see Section~\ref{sec:sec4_2}).
In this model, X-rays are produced by the synchrotron from shock-accelerated electrons.
The {\it H.E.S.S.} emission is then produced by external Compton (EC)
and synchrotron-self Compton (SSC) up-scattering of the stellar photons and
synchrotron photons. In the emission zone, the photon density of the synchrotron emission is
$\sim$10 times larger than that of the blackbody emission, which causes SSC to dominate at the
highest energies.
The {\it Fermi}-LAT data are hard to explain with the simple synchro-Compton
model. In stochastic shock acceleration theories, the maximum electron energy is limited by the
radiation reaction, and the shock accelerated electrons cannot emit synchrotron photons
above $\sim$160\,MeV. Even with bulk acceleration in the shock (see Section~\ref{sec:sec4}),
it is hard to explain the full {\it Fermi}-LAT data. Furthermore, if the {\it Fermi}-LAT photons are
produced by the synchrotron process, the light curve should correlate well with the X-ray light
curve. However, the high-energy {\it Fermi}-LAT light curve does not correlate with the X-ray
light curve, hence there must be some other processes responsible for the {\it Fermi}-LAT photons.
We attribute this to pulsar magnetosphere emission at a few GeV and to EC at the highest energies
as we discuss below (Section~\ref{sec:sec4}).

\subsection{Gamma-ray variability}
\label{sec:sec3_5}
           We first checked for secular variation of the gamma-ray flux. 
This is particularly  interesting as the X-ray flux at phase 0 (the spike)
is seen to be highly variable, and the other phases are relatively stable.
The 3FGL variability index \citep[][]{fermi3fgl} is 42, implying no significant phase-averaged
variability on a month timescale. Here we investigate variability on shorter time scales, $1-16$\,days.

             We divided the observation into individual orbits (16.544-day interval).
For each time interval, we performed a likelihood fit while holding
all the parameters fixed at the mission-averaged values
for the corresponding phase (Section~\ref{sec:sec3_2})
except for the J1018's normalization. We measured the source flux for each time interval
and constructed the light curve. We then calculated $\chi^2$ for a constant flux
for the time intervals where the fit successfully converged and the TS value
for J1018 is greater than 1. We also varied the model to include other nearby
bright sources and/or to let the J1018 spectral shape parameters vary,
and find that the best fit was obtained when freeing only the J1018's normalization.
In this case, the $\chi^2$/dof for a constant flux is 173.7/142, implying 3\% chance
that the source flux is constant in time. 

We next attempted to study the variability in phase, binning into 1.65-day intervals
and then assembling light curves for each of these 10 orbital phase bins.
We find no significant ($3\sigma$) flux variability in any phase bins, although
the chance probability at phase 9 is relatively low ($p=3$\%).
Combining phase bins 0 and 9 yields a similar low ($p=3$\%) variability significance.

\section{Emission modeling}
\label{sec:sec4}
        There is growing evidence that the compact object in J1018 is a
neutron star, although no study is yet conclusive \citep[][]{wr15,abb+15,scc15,wrk+15}.
Below we model the emission from J1018 under the assumption that the compact object
is a neutron star.

        If the compact object in J1018 is an energetic spin-powered neutron star,
there should be winds from both the neutron star and the optical companion.
The two winds form a contact discontinuity (CD) where the ram pressures of the two winds balance.
If the stellar wind momentum flux is larger than that of the pulsar wind,
the CD curves towards the pulsar. Particles are accelerated to high energy
in the shocked pulsar wind. We additionally expect some adiabatic acceleration
 \citep[e.g.,][]{bkk08,dlf15}
as this shocked wind flows away from the apex along the CD. This is enhanced by the increasingly
tangential orientation of the pulsar wind shock as one moves downstream. The
net effect is a growing bulk Lorentz factor ($\Gamma$) for the radiating shocked wind.
The particles in the shock emit
photons via synchrotron and inverse-Compton processes (synchro-Compton model).
In this picture the X-ray synchrotron emission is strong at two orbital phases:
the periastron and the inferior conjunction of the compact object. This may be
able to explain the peculiar X-ray light curve of J1018 \citep[][]{abb+15}.

        The observed {\it Fermi}-LAT light curves (Figure~\ref{fig:fig3}) suggest that there should be
at least two emission components in that band as noted above (Section~\ref{sec:sec3_2}).
Furthermore, the substantial steady component in the {\it Fermi}-LAT band suggests
the existence of an additional ``DC'' component which we attribute to the pulsar
magnetosphere. Keeping these in mind, we model the light curve and the SED of J1018 below. 

\subsection{Light curve modeling}
\label{sec:sec4_1}

\begin{figure*}
\centering
\begin{tabular}{cc}
\hspace{10.0 mm}
\includegraphics[width=2.7 in]{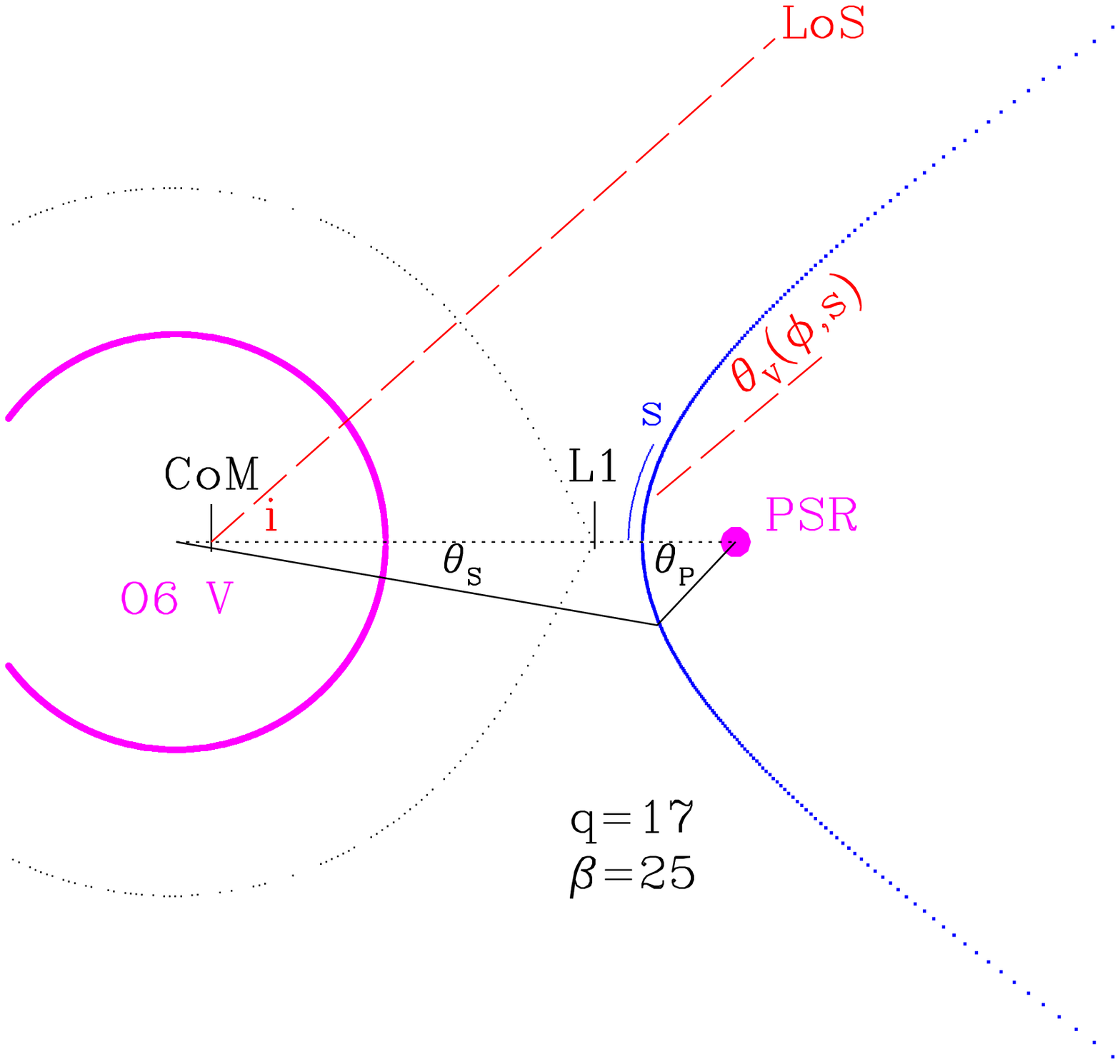} &
\hspace{0.0 mm}
\includegraphics[width=4.0 in]{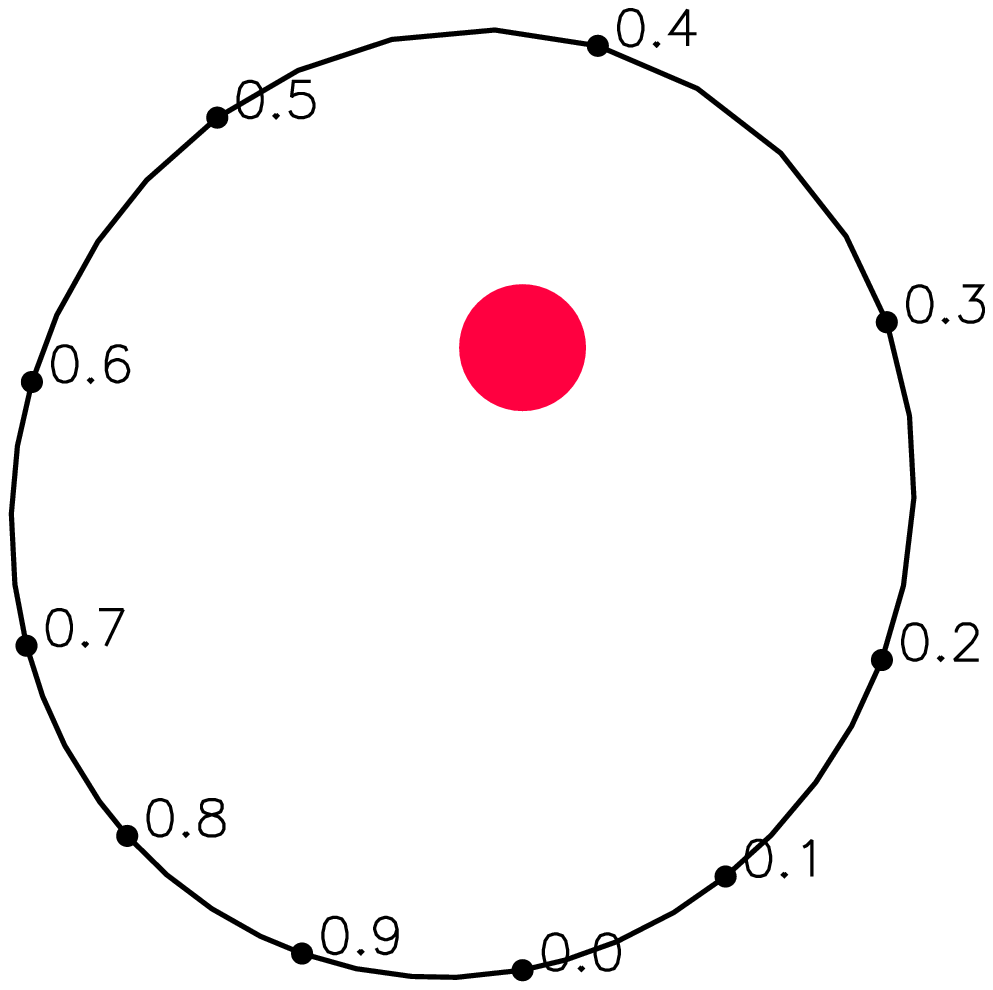} \\
\end{tabular}
\figcaption{Left: A vertical cross section of the system. Various
geometrical parameters are denoted.
Right: A horizontal cross section of the orbit.
The position of the neutron star (black dots) in the orbit is shown as a function
of orbital phase $\phi_{\rm orb}$. The observer looks at the system from
the bottom of the page at an inclination of 50$^\circ$. 
\label{fig:figX}
}
\vspace{0mm}
\end{figure*}

        We model the X-ray light curve with the synchrotron emission produced
by particles flowing along the CD.
Near the apex of the CD, the shocked wind has small $\Gamma$ and so
emits a nearly isotropic component whose strength varies with the particle and field energy
density variations due to the varying IBS standoff distance in the eccentric orbit.
The isotropic emission is $F_{\rm iso} \propto B^2 \propto r_{\rm psr}^{-2}$
(for the assumed synchrotron radiation and a transverse $B$ in a stripped wind), where $r_{\rm psr}$ is
the distance from the pulsar to the apex of the CD. The distance is given by the
ram pressure balance and is
$$r_{\rm psr}=\frac{a(1-e)}{(1 + e\mathrm{\ cos}\phi)}\frac{1}{(1+\sqrt{\eta})},$$
where $a$ is the semi-major axis, $e$ is the eccentricity, $\phi$ is
the phase angle (true anomaly), and 
$\eta$ is the momentum flux ratio of the stellar and pulsar winds.
We assume that the masses of the pulsar and the companion are
$1.4M_{\odot}$ and $23M_{\odot}$, that the winds of both
stars are isotropic and that the wind momentum flux ratio is $\eta=25$.
We choose these parameters to be roughly consistent with previous studies \citep[][]{scc15,d06}.

\newcommand{\marka}{\tablenotemark{a}}
\begin{table}[]
\vspace{-0.0in}
\begin{center}
\caption{Parameters used for the light curve and the SED models.
\label{ta:ta2}}
\vspace{-0.05in}
\scriptsize{
\begin{tabular}{ccccc} \hline\hline
Parameter           & Symbol             & Value  \\ \hline
Eccentricity        & $e$                & 0.35    \\
Inclination (deg.)  & $i$                & 50      \\ 
Semi-major axis (cm)& $a$                & $5.5\times10^{12}$      \\ 
Momentum flux ratio & $\eta$             & 25    \\
Max. bulk Lorentz factor & $\Gamma_{\rm max}$ & 7       \\ 
Magnetic field strength (G) &  $B$ & 1.5  \\ \hline
Low-energy spectral index & $p_{\rm 1}$ & $1.93$  \\
High-energy spectral index & $p_{\rm 2}$ & 2.15  \\
Minimum electron energy & $\gamma_{\rm e,min}$ & $5\times10^{3}$  \\
Maximum electron energy & $\gamma_{\rm e,max}$ & $10^{8}$\marka       \\
Break electron energy   & $\gamma_{\rm e,b}$ & $5\times10^{7}$\marka \\
Injected particle energy ($\rm erg\ s^{-1}$) & $\dot E_{\rm P}$ & $6\times10^{35}$  \\
Injected magnetic energy ($\rm erg\ s^{-1}$) & $\dot E_{\rm B}$ & $3\times10^{34}$ \\ \hline
\end{tabular}}
\end{center}
\hspace{-2.0 mm}
$^{\rm a}${Varies along the shock.}\\
\end{table}

        The shape of the CD is calculated following \citet{crw96} (see Figure~\ref{fig:figX} left).
Note that, with a synchrotron cooling time $\tau_c \approx \frac{3m^3c^5}{2e^4\gamma B^2}
\approx 5 \times 10^8 \gamma^{-1}B^{-2}$s$\approx 2$s (for Table 2 parameters) 
shorter than the characteristic flow time along the contact discontinuity 
$\tau_f \approx a/\beta c \approx 40/\beta$s, the high energy electrons can cool 
in the `slow population', placing a spectral break in the $\sim$MeV
range (Figure~\ref{fig:fig2}). The lower energy X-ray photons emitted at different
distances $s$ along the shock from the apex will have the same spectral
slope as that of the isotropically emitted photons near the apex ($\Gamma \approx 1$).
Hence Doppler boosting amplification provides the principal variation in emissivity along the shock.
Thus for a distance $s$ from the apex where the bulk Lorentz factor has grown
to $\Gamma(s)$ the synchrotron emission viewed at angle $\theta_{\rm v}(s)$ is
\citep[e.g., see equation 3 of][]{fdb08}:
$$\mathcal{A}(s) \propto N_{\rm e}(s)\delta_{\rm D}^{(5+p_{\rm 1})/2}(s,\theta_{\rm v})B^{(1+p_{\rm 1})/2}(s),$$
where $N_{\rm e}$ is the number of electrons, 
$p_{\rm 1}$ is the spectral index of the power-law electron distribution,
$\delta_{\rm D}=[\Gamma(1 - \sqrt{1 - 1/\Gamma^2}\mathrm{cos}\theta_{\rm v})]^{-1},$ is 
the Doppler factor, and $B(s)$ is the shock compressed magnetic field strength.
Here, we assume that $\Gamma \propto s$, the cumulative shocked plasma
$N_{\rm e}\propto 1 - \mathrm{cos}\theta_{\rm p}$ where
$\theta_{\rm p}$ is the angle between the vector to the emission region from the pulsar
and the line of centers, and $B$ is proportional to the distance
from the pulsar and the emitting region in the flow
($B\propto r_{\rm p}^{-1}$, where $r_{\rm p}$ is the distance between the pulsar
and the emission region).

	We are identifying the spike with the Doppler-boosted emission beamed along the
IBS. Since this spike is narrow, we require a moderately large
$\delta_{\rm D} \gapp 3$. However the spike, while highly variable, is typically only
2--3$\times$ brighter than the phase-averaged flux. Since for $p_1 \approx 2$ the 
amplitude scales as $\sim\delta_{\rm D}^{3.5}$, only a small factor ($<10\%$) of
the shocked wind can be so highly boosted. Thus we envision two components in
the shocked pulsar wind -- a low $\sim$constant $\Gamma$ dominant zone that flows 
along the CD and a higher $\Gamma \propto s$ skin due to adiabatic acceleration and
the increasingly tangential nature of the pulsar wind shock. Indeed hydrodynamic simulations
\citep[][]{bkk08, dlf15} do find such complex post-shock flow patterns.

\begin{figure*}
\begin{tabular}{cc}
\hspace{-3.0 mm}
\includegraphics[width=3.3 in,angle=0]{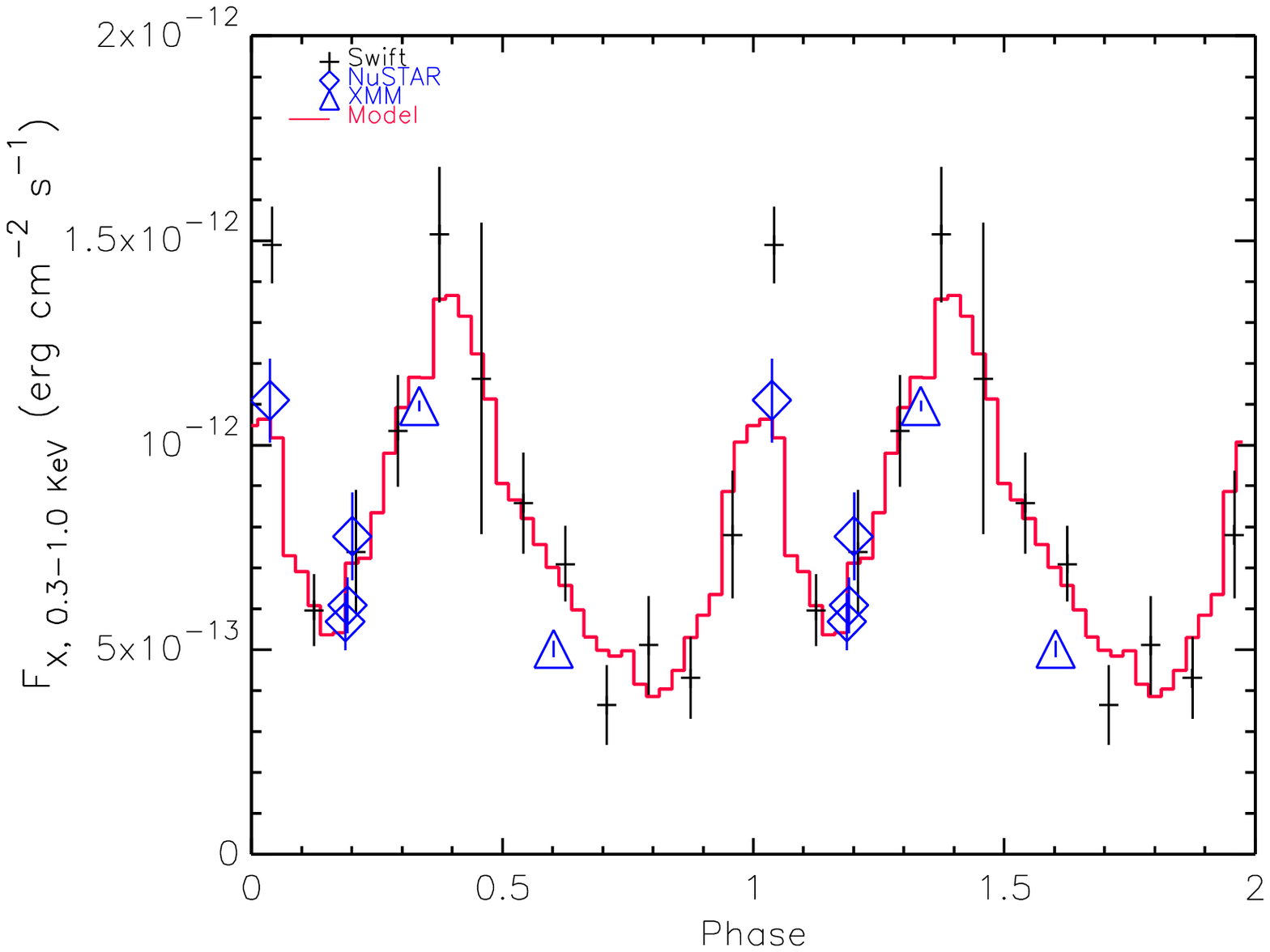} &
\hspace{-5.0 mm}
\includegraphics[width=3.25 in]{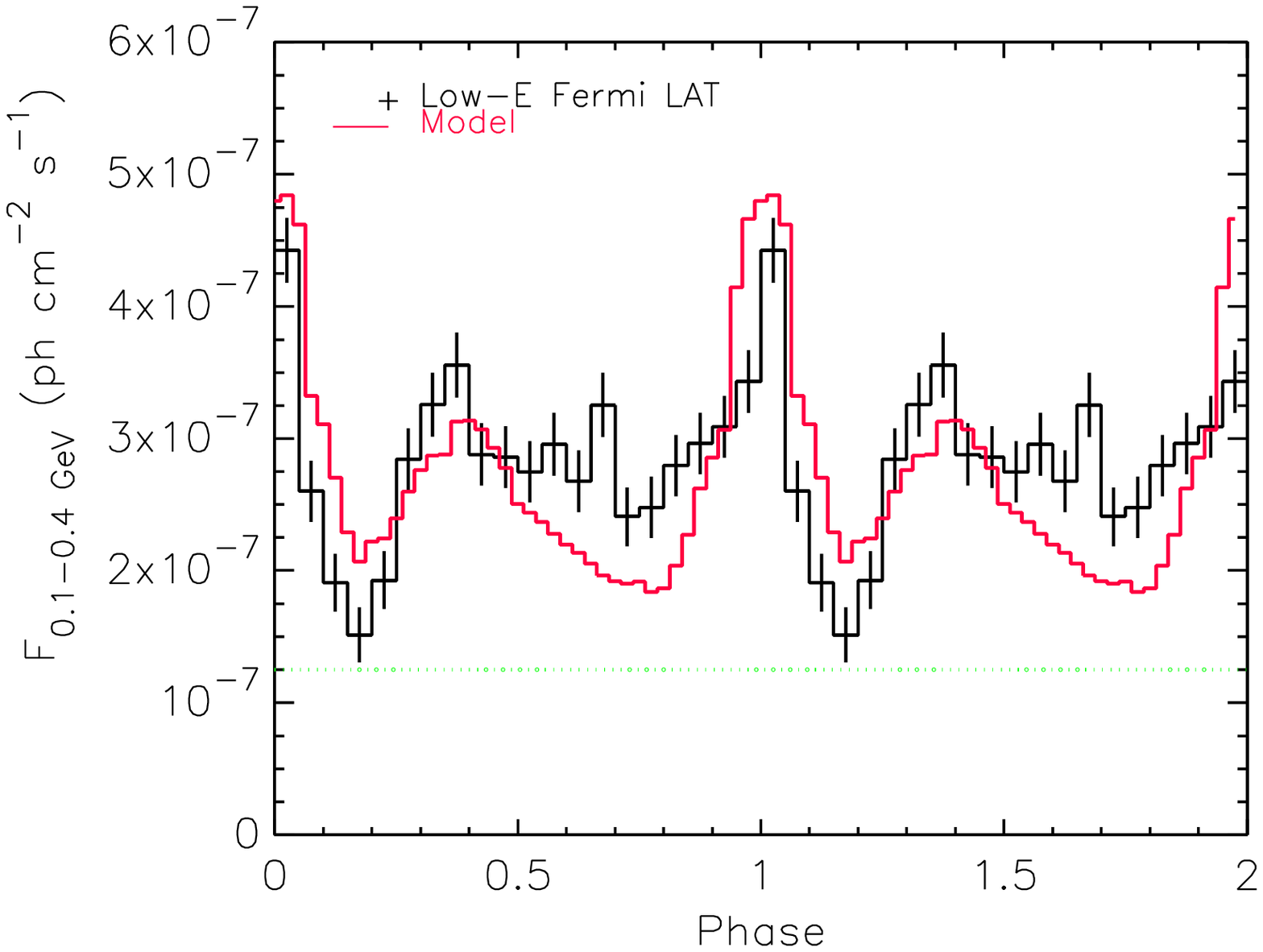} \\
\hspace{-3.0 mm}
\includegraphics[width=3.25 in,angle=0]{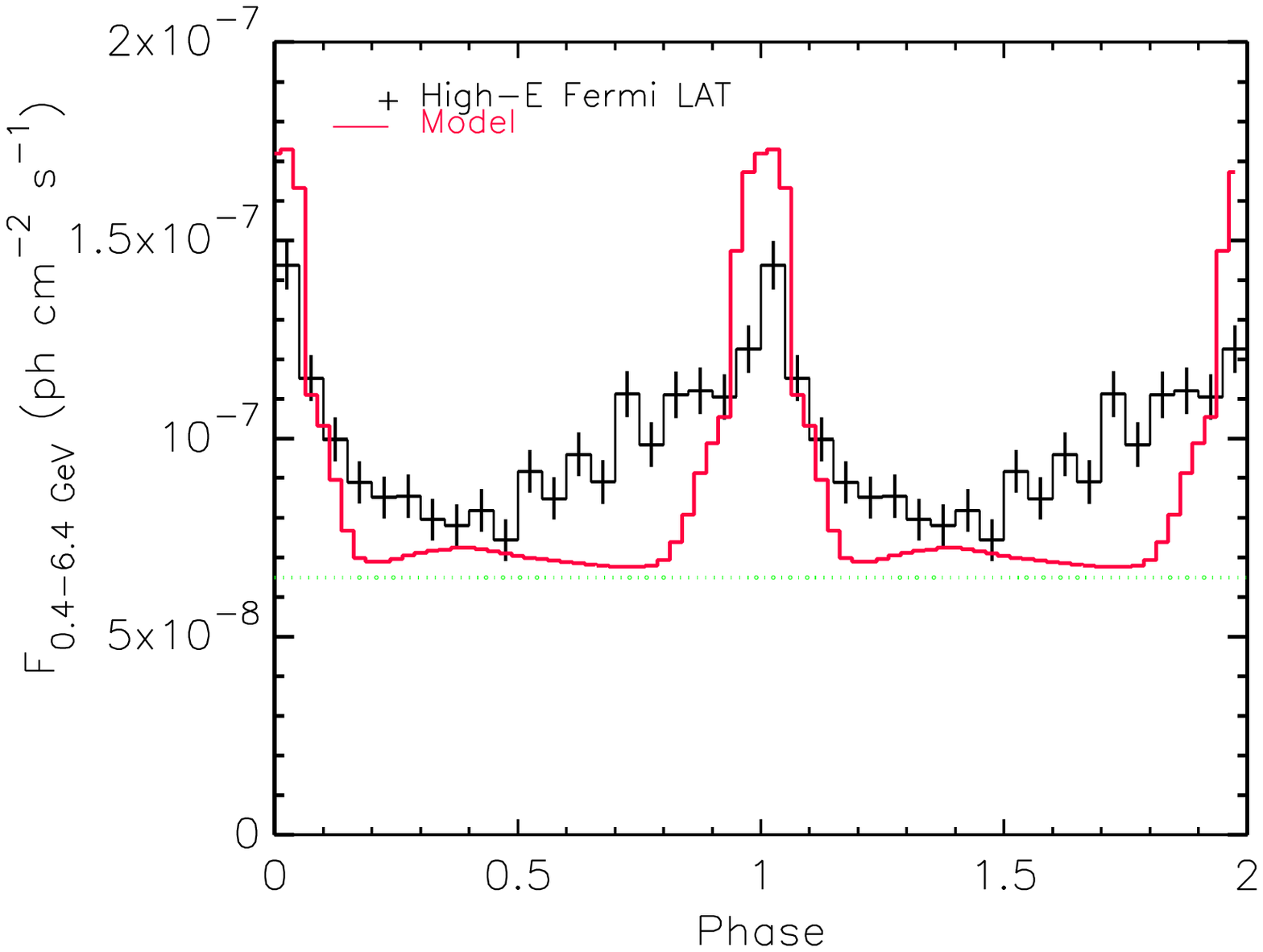} &
\hspace{-5.0 mm}
\includegraphics[width=3.1 in]{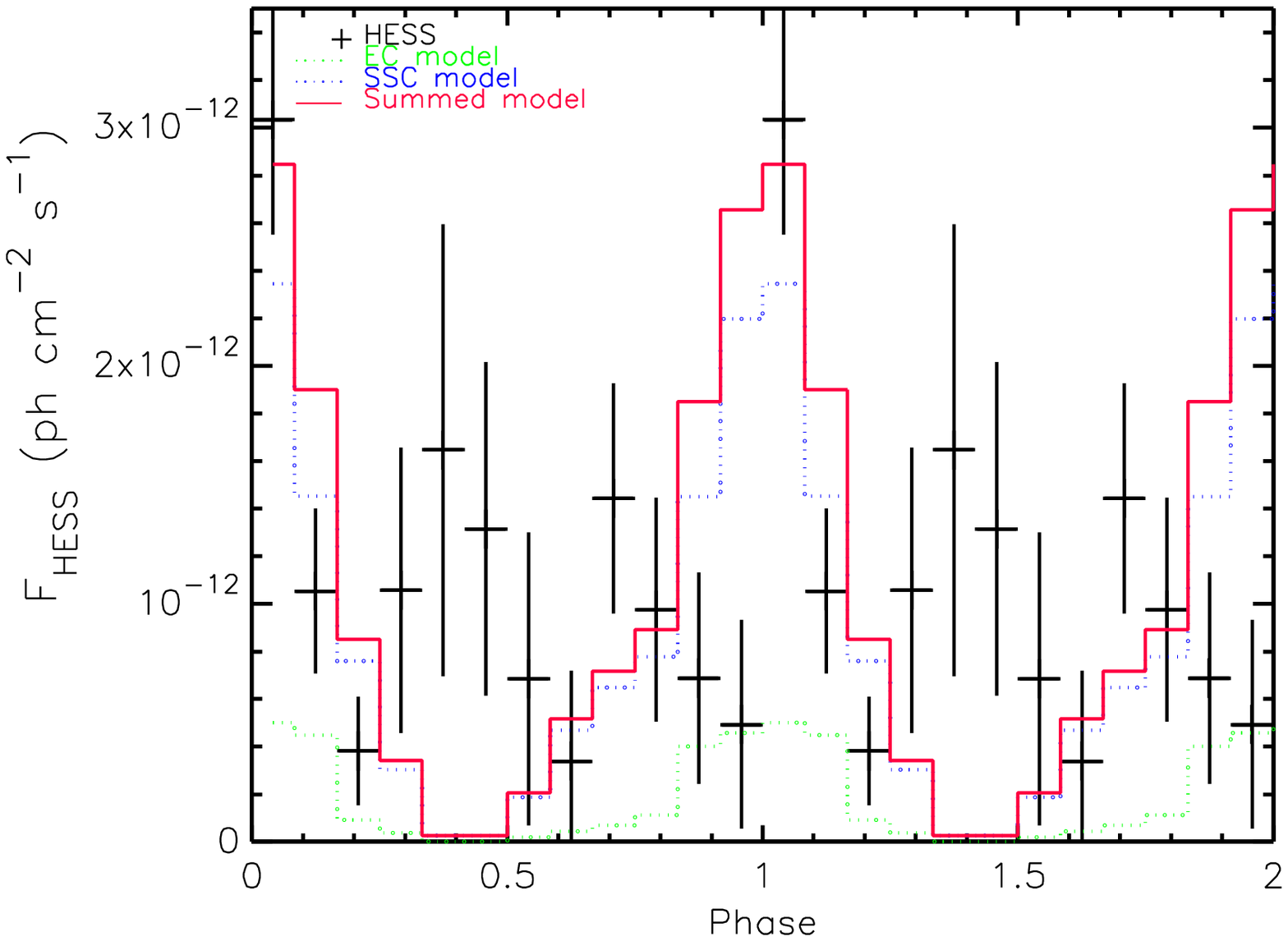} \\
\end{tabular}
\vspace{0.0 mm}
\figcaption{
{\it Top left}: The X-ray light curve and a light curve model.
The high flare state seen with {\it Swift} is not shown
in this plot.
{\it Top right}: Low-energy (100--400\,MeV) {\it Fermi}-LAT light curve and a light curve model.
{\it Bottom left}: High-energy (0.4--6.4\,GeV) {\it Fermi}-LAT light curve and a light curve model.
The green line in the top-right and bottom-left panels shows the pulsar contribution.
{\it Bottom right}: The {\it H.E.S.S.} light curve and a light curve model.
The {\it H.E.S.S.} data of \citet{J1018hess15} is newly folded at the X-ray orbital period
used in this paper (de O{\~ n}a-Wilhelmi, private communication).
The parameters for the model are $i=50^\circ$, $P_{*}c/E_{\rm psr}=25$, and $\Gamma_{\rm max}=7$
(see Section~\ref{sec:sec3_2}).
\label{fig:fig4}
}
\vspace{0mm}
\end{figure*}

	To summarize, we have a low $\Gamma$ component modulated by the orbital
geometry and a higher $\Gamma$ component whose observed intensity is primarily 
controlled by the line-of-sight beaming angle. The former will be strongest at
periastron, which we take to be $\phi_{\rm orb}=0.39$
so that the peak of the model hump coincides with the maximum of the X-ray hump in phase.
The latter will be strongest at $\phi_{\rm orb}\approx 0$ generating the spike
(see Figure~\ref{fig:figX} right and Figure~\ref{fig:fig4} top left).

         We assume that a broken power-law electron population with a low-energy index
of 1.93 so that the model goes through the phase-variable
X-ray spectrum (Figure~\ref{fig:fig2}),
and a high-energy index of 2.15 is injected at the shock (see Section~\ref{sec:sec4_2}).
Note that the light curve model is insensitive to the exact spectral index.
For each phase angle $\phi_{\rm orb}$ in the orbit of the binary, we consider
the synchrotron emission of the flows towards the observer at $\phi_{\rm orb}=0^\circ$ 
(corresponding to 201$^\circ$ from the periastron) and inclination $i$ 
and integrate over the shock surface, covering a range $s=0-3\times$ the orbital separation.
We find that this two-component flow explains the X-ray light curve with
3\% of the flow in 
the accelerated component. Figure~\ref{fig:fig4} top left shows the computed X-ray light 
curve model compared with the data. 
The eccentricity is constrained by the shape of the sinusoidal hump, and the inclination
is by the shape and amplitude of the spike -- note in particular that we have chosen $i$ so that
the Earth line-of-sight (LoS) is close to grazing incidence for the IBS.
At smaller $i$ the spike will be absent and
at larger $i$ the spike will be double.
The parameters used for the model shown in the figure are $e=0.35$, $i=50^\circ$,
$B_{\rm apex}=1.5$\,G, and maximum $\Gamma=7$ (see Table~\ref{ta:ta2}).
Note that simple power-law extension of the injection spectrum to lower energies does
not alter the fit, hence $\gamma_{\rm e, min}$ is not well constrained.

  For the same model we can compute the Sy/EC/SSC components and compare with 
gamma-ray light curves (Figure~\ref{fig:fig4}).
At the top right we show the model and data for the low energy {\it Fermi}-LAT band. Here
synchrotron emission from above the break dominates. For simplicity we have assumed
that the injection spectrum does not vary around the orbit and so the shape follows that
of the X-ray light curve, with isotropic (hump) and beamed (spike) components.
Notice that the rather narrow minimum at $\phi_{\rm orb}\approx 0.2$ and the detailed
spiky shape of the hump are not well matched; we comment on this in Section~\ref{sec:sec5}.
Moving to the $>0.4$\,GeV {\it Fermi}-LAT band, we are above the spectral cut-off of the
unboosted (hump) population. Indeed only the boosted (spike) populations contributes.
In both {\it Fermi}-LAT panels a phase-independent pulsar contribution (green line) is included.
The true amplitude is somewhat uncertain.
Moving to the TeV band (bottom right), we expect EC/SSC to dominate. Note that the {\it H.E.S.S.}
light curve is dominated by the $\phi_{\rm orb}=0$ emission\footnote{This light curve
is folded on the refined orbital period and
is slightly different from that reported by \citet{J1018hess15}.}.
At first sight this might be surprising,
since the EC emission (upscatter of stellar photons) should peak at periastron (broad hump, 
$\phi_{\rm orb}=0.39$).  However, photon-photon absorption \citep[][]{gs67,d06b}
by the stellar photons strongly suppresses this component (and the SSC emission from this phase).
The absorbed EC and SSC components are shown by the green and blue lines and the total
Compton emission shows the phase 0 spike. The slight over-absorption
seen at $\phi=0.3-0.5$ may be due to our approximate IBS geometry; some downstream
SSC photons can be less strongly absorbed.

\subsection{Phase-resolved SED model}
\label{sec:sec4_2}

\begin{figure*}
\centering
\hspace{0.0 mm}
\includegraphics[width=7. in]{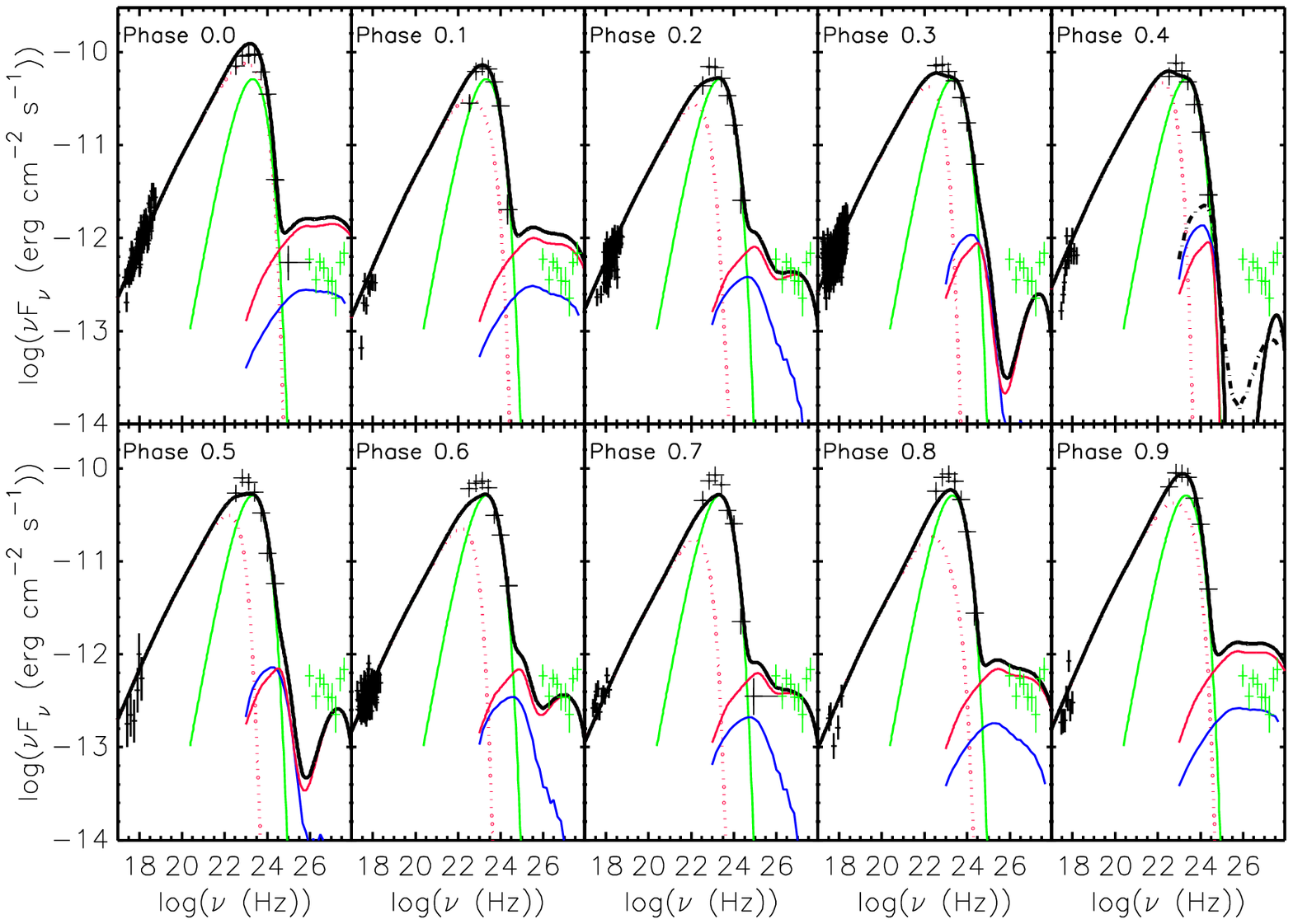} \\
\figcaption{Phase-resolved broadband SEDs and the emission model (Section~\ref{sec:sec4_2}).
The X-ray-to-{\it Fermi}-LAT data are phase-resolved (black crosses), and
phase-averaged {\it H.E.S.S.} data \citep[green crosses;][]{J1018hess15} are also shown for reference.
The model is calculated for 40 phase bins and re-summed to the 10 panels shown.
Lines are for model components: synchrotron (red dotted),
pulsar (green), SSC (red solid), EC (blue solid), and the sum (black solid).
Photon-photon absorption is calculated assuming that emission is
all from the apex of the shock. However, for phase 0.4, we also show the high-energy SED with photon-photon
absorption calculated using the realistic shock geometry (black dot-dashed line).
See Section~\ref{sec:sec4_2} and Table~\ref{ta:ta2} for the model and the parameters.
Notice that the synchrotron emission goes up to the highest frequency at phase 0 as expected
due to the high bulk Lorentz factor.
\vspace{0.0 mm}
\label{fig:fig5}
}
\vspace{0mm}
\end{figure*}
 
          We model the SED with a synchro-Compton
model with an addition of a pulsar magnetosphere component.
This pulsar emission is a power law with sub-exponential cutoff,
$dN/dE=N_{\rm 0} (E/E_{\rm 0})^{-\Gamma_{\rm psr}}{\rm exp}[-(E/E_{\rm c})^b]$,
where we used $N_{\rm 0}=5.7\times10^{-7}\rm\ ph\ cm^{-2}\ s^{-1}\ GeV^{-1}$,
$E_{\rm 0}=1$\,GeV, $\Gamma_{\rm psr}=0.7$,
$E_{\rm c}=110$\,MeV,
and $b=0.48$, corresponding to
$L_{\gamma}\simeq5\times10^{35}\rm\ erg\ s^{-1}$ for
an assumed distance of 5\,kpc to match the observed {\it Fermi}-LAT SED.
Note that these pulsar parameters are typical for
gamma-ray pulsars with sub-exponential cutoff spectra \citep[][]{fermi2PC}.
The star emits blackbody photons with $kT\sim4$\,eV and $L_{\rm BB}=10^{39}\ \rm erg\ s^{-1}$
and provides seed photons for EC scattering, and absorbs the TeV emission.

            We use the same parameters that we used for the X-ray light curve
modeling (see Section~\ref{sec:sec4_1}) and compute the EC and SSC emission.
Note that we use a broken power-law electron distribution.
A broken power-law distribution can be formed due to particle cooling \citep[e.g.,][]{msca05}.
In our case, electrons are injected by the pulsar over the whole shock. Those injected near the apex
flow and cool along the shock. Therefore, at any point in the shock, there are cooled electrons
and freshly injected electrons. This process necessarily produces a broken power-law distribution.
Note that the break $p_{\rm 2}-p_{\rm 1}$ is not 1 in our case; a break of 1 is expected
for electrons cooling in homogeneous magnetic field. In our model, the magnetic field
strength decreases with distance ($s$), hence the break may be smaller.

         We adjust the maximum electron Lorentz factor ($\gamma_{\rm e,max}$) at each
point of the shock so that the maximum synchrotron frequency in the rest frame of the flow
is $\sim$160\,MeV, the radiation reaction limit. The spectral break energy evolution is
calculated by integrating $\dot \gamma \propto -B^2 \gamma^2$, assuming synchrotron cooling
dominates. We also compute the gamma-gamma absorption between the high-energy EC/SSC photons and the
stellar photons \citep[see, for example,][]{d06b}.
For this, we assumed that the optical star is a point source and all the high-energy TeV photons are
emitted at the apex of the shock for simplicity. We calculate the differential optical depth
given in \citet{d06b} at each point along the LoS and integrate it out
to $\sim$10$\times$ the distance between the star and the emitting region. 
Beyond this distance, the blackbody photon field becomes negligibly weak,
as does the absorption. The absorption is then computed by taking the exponential of the
integrated optical depth.

Note that here we calculated the effect of absorption only for spectrum emitted
at the apex of the shock in each phase and assumed that this is the same for
emission over the whole shock.
The absorption is strongest at phase $\sim$0.4 (when the compact object is
behind the optical companion) and at $\sim10^{25}$\,Hz. Hence the effect of
the absorption is to remove the sinusoidal hump in the (low-energy)
{\it H.E.S.S.} light curve. We calculate the model emission for 40 phase bins and show
the phase-averaged SED and the model in Figure~\ref{fig:fig2}. We note that the model
also explains the phase-resolved SEDs reasonably well (Figure~\ref{fig:fig5}).

         The maximum photon frequency of the synchrotron emission
is highest at phase 0 because of Doppler boosting.
Hence, a significant amount of high-energy synchrotron photons ($\gapp 1$\,GeV)
is produced only near phase 0, and the pulsar emission dominates the GeV band in the other phases.
Photons in the sinusoidally modulating phases extend only up to $\sim 200$\,MeV (no Doppler boosting),
hence are visible only in the lower-energy {\it Fermi}-LAT light curve.
Therefore, the low-energy {\it Fermi}-LAT light curve shows two peaks while the high-energy one
has only one (Figure~\ref{fig:fig4}).

\section{Discussion and Conclusions}
\label{sec:sec5}

	Gamma-ray binaries such as LS~5039 show only broad variation in the TeV
band, which is interpreted as orbital modulation of EC emission. J1018 uniquely
shows two components in the {\it Fermi}-LAT band, a broad modulation and a sharp spike;
the spike continues into the TeV range.
Moreover the finding of \citet{scc15} that the spike phase corresponds to compact
object (neutron star) inferior conjunction argues against a simple EC interpretation.
We have shown that this multi-component picture can be understood by positing slow
and faster (moderate $\Gamma$) flow in a shocked pulsar wind. Both components
contribute synchrotron emission in the X-ray to the low-energy {\it Fermi}-LAT band; only the 
Doppler boosted component reaches energies above 1\,GeV. The broad hump (low $\Gamma$) emission
is very asymmetric.  In our model, this is explained by orbital eccentricity ($e\sim0.35$).

	The spike emission depends on Doppler boosting and thus places significant 
constraints on the $\Gamma$ and viewing geometry.
Existence of such beamed emission in gamma-ray binaries has been suggested \citep[e.g.,][]{dlf15,d15},
and similar phenomena have been seen in some pulsar binaries \citep[e.g.,][]{rs16} as well.
Modeling the spike allows us to infer the properties of the fast flow and
the orbital parameters; the maximum bulk Lorentz factor of the flow is $\sim$7
and the inclination is $\sim$50$^\circ$ (for $\eta=25$).
This inclination is in agreement with estimates obtained by spectroscopic study 
\citep[$32^\circ-49^\circ$ for $M_{*}=20M_{\odot}$ or
$39^\circ-64^\circ$ for $M_{*}=26.4M_{\odot}$;][]{scc15}.
It should be noted that the required inclination is somewhat dependent on the 
momentum flux ratio $\eta$ assumed here, and might be somewhat different in
a detailed hydrodynamic simulation \citep[e.g.,][]{bkk08}.

           Our interpretation on the X-ray spike enables several specific predictions.
First, because the beamed emission is produced by synchrotron radiation of cooled electrons,
the spectrum should have a different spectral break at different phases. While we expect
that the actual break energy is inaccessible at $\sim$MeV energies, precision X-ray
phase spectra might be compared with phase-resolved {\it Fermi}-LAT spectra at low energies
to reveal break shifts.
Also, the X-ray flux variability of the spike is expected because a small change of the
beaming angle (shape of the CD) due to the variation of stellar wind momentum flux can result in a
large change in the flux. Our current constraint on the inclination is such that the
observer sees the shock slightly above the asymptotic tangent of the shock.
Larger stellar winds will make the opening angle of the shock smaller and the synchrotron
flux will drop.  In principle a large drop in the stellar wind flux might allow the
IBS to expand sufficiently to show a broader or even doubled peak bracketing phase 0.
This can be tested with optical/X-ray monitoring to infer $\eta$ variations.

                   The change of the light-curve shape in the {\it Fermi}-LAT band suggests
another constant emission component in that band. We assume that this component
is the pulsar magnetosphere emission. Alternatively,
it could be EC up-scattering of the stellar photons
by a lower-energy relativistic Maxwellian distribution as hypothesized by \citet{dlf15}.
If so, this component will produce a broad hump peaking at the periastron in the light curve.
Perhaps, some of the low-energy {\it Fermi}-LAT flux may be attributed to this.
Whether or not this can explain the constant flux is unclear.
Nevertheless, if our interpretation of the {\it Fermi}-LAT-band emission is correct
(i.e., the pulsar contributes significantly),
it may be possible to detect gamma-ray pulsations when the orbital parameters are better
constrained.

            Our model captures the main features of the {\it Fermi}-LAT light curves (Figure~\ref{fig:fig4}).
In particular, boosted synchrotron
emission accounts for the spike at phase 0 and contributes to the
bump at phase 0.3--0.4. However, the bump is too weak and we
overproduce the emission at phase 0.1--0.2. We cannot attribute the
dip at this phase to absorption since the pulsar is in front. Thus
we infer a somewhat lower pulsar contribution at these energies and
a larger modulation of the synchrotron shock flux.

        A possible source of such modulation is asymmetry in the pulsar
wind; many young pulsars have wind nebulae concentrated in an equatorial
torus. If such a torus is inclined with respect to the orbital plane then the two
cross at two phases. Thus we can imagine the pulsar wind pole pointing
to phase $\phi \sim 0.1-0.2$ with a weaker shock and decreased synchrotron
emission, while the equatorial flow at phase $\phi \sim 0.35-0.5$ might
give rise to enhanced emission, near periastron. At phase 0.8--0.9
the wind interaction would be weakened by the larger orbital separation.
Note that the enhanced synchrotron emission could add additional SSC flux
at $\phi \sim 0.35-0.5$, where the {\it H.E.S.S.} data also lie above the model.
Modeling such an anisotropic wind lies beyond the scope of this paper.

        In the TeV band we see boosted SSC at the phase 0 spike. This
appears especially sharp in the refolded {\it H.E.S.S.} data and so some amendments
to the model (e.g. sharper shock spatial curvature due to an anisotropic
pulsar wind and/or higher bulk $\Gamma_{\rm max}$) might be useful.

	We can also
expect additional SCC modulation if the pulsar wind is anisotropic.
We generally expect a broad sinusoidal hump from EC emission in the
$\sim 10$\,GeV--TeV band. However in our model absorption from the soft
stellar photons nearly completely suppress this component (Figure~\ref{fig:fig5}).
Our model computes a simple exponential attenuation of the shock flux,
assuming that the emission is located near the apex and integrating the
optical depth of the stellar photon field along the instantaneous
line of sight. This is almost certainly an over-estimate.

        To test this, for phase 0.4 we recomputed the absorption by
computing the optical depth for each emission point along the shock
surface. As the distance $s$ from the apex increases, absorption decreases
for that portion of the shock above the orbital plane and increases for
the part of the shock that lies below. Given the exponential nature
of the attenuation, the effect on the total emission integrated over the
entire surface is a decreased absorption, up to $6\times$ less
than in the point source approximation. This estimate is shown as the dashed
line in Figure~\ref{fig:fig5}. The details of this absorption depend on the detailed
shock structure (e.g. modified by anisotropic pulsar emission), so further
analysis can await more detailed observation.

        A second effect can also boost the observed TeV flux at these
phases: secondary $e^+/e^-$ pairs produced in TeV absorption on stellar
photons can emit in the $\sim 10^{24} - 10^{25}$\,Hz band \citep[e.g.,][]{b13,d13}.
This cascade emission can be especially
important near pulsar superior conjunction and so may also contribute to
the lower energy TeV emission at $\phi \sim 0.3-0.5$. TeV phase-resolved
spectra anticipated in the Cherenkov Telescope Array (CTA) era would
certainly motivate such detailed modeling of the absorption and re-emission.

	In addition to reproducing the multiband light curves fairly well, our model 
can match the phase-resolved SEDs (Figure~\ref{fig:fig5}). The gamma-ray luminosity of
the putative pulsar is $\sim$$10^{36}\rm \ erg\ s^{-1}$. This suggests a very energetic young
pulsar (consistent with the young massive binary). Comparing with \citet{fermi2PC}
suggests a parent pulsar with $\mathrm{Log}({\dot E})\approx 36-39$, which could certainly 
supply the required particles and magnetic energy for the shock (Table~\ref{ta:ta2}).
A gamma-ray pulse detection would be a crucial input to a more detailed model.

            The {\it H.E.S.S.} spectrum appears to rise above 10\,TeV. This is not
a feature of our current SED, but could be easily accommodated if we introduced
low-energy seed photons ($\nu\lapp10^{12}-10^{13}$\,Hz) and maintained a hard
$e^+/e^-$ injection. While radio-millimeter observations can probe this population,
it may be difficult to compute the $\gapp$10\,TeV spectrum since Compton-upscattering
of these soft photons depends on the poorly known
scattering geometry. Much of the population
of these slow-cooling electrons may be at large $s$ and less useful for Compton
upscatter. An inner-system low-energy synchrotron component however might
be due to low-energy pulsar injection as hypothesized by \citet{dlf15}.
Another possible contribution to the TeV flux might be traced to the very
hard (albeit with large uncertainty) X-ray spectra seen by \citet{adkh13} at 
several phases. These data may require the injection spectrum to vary with orbital
phase. This could be another signature of an anisotropic pulsar wind and we would
benefit from better constraints on the X-ray spectrum phase variation. Detailed
phase information on the $\gapp$10\,TeV flux would also be quite helpful -- and should
be available in the CTA era.

         We conclude that 1FGL~J1018.6$-$5856
can provide a good opportunity to study
shock acceleration and hydrodynamic flow in gamma-ray binaries thanks to its peculiar
X-ray-to-gamma-ray light curves, with multiple emission components appearing
in the several wavebands. While our observationally driven model reproduces
important features of the light curves and SEDs, detailed relativistic hydrodynamic
simulations and monitoring observations may well be needed for a full understanding.
As a test case with particularly rich behavior, J1018 can certainly give us new 
insights into IBS emission in gamma-ray binaries.
\bigskip

We thank E. de O{\~ n}a Wilhelmi and the {\it H.E.S.S.} collaboration for
providing the {\it H.E.S.S.} light curve folded on the refined orbital period.
The \textit{Fermi} LAT Collaboration acknowledges generous ongoing support
from a number of agencies and institutes that have supported both the
development and the operation of the LAT as well as scientific data analysis.
These include the National Aeronautics and Space Administration and the
Department of Energy in the United States, the Commissariat \`a l'Energie Atomique
and the Centre National de la Recherche Scientifique / Institut National de Physique
Nucl\'eaire et de Physique des Particules in France, the Agenzia Spaziale Italiana
and the Istituto Nazionale di Fisica Nucleare in Italy, the Ministry of Education,
Culture, Sports, Science and Technology (MEXT), High Energy Accelerator Research
Organization (KEK) and Japan Aerospace Exploration Agency (JAXA) in Japan, and
the K.~A.~Wallenberg Foundation, the Swedish Research Council and the
Swedish National Space Board in Sweden.

Additional support for science analysis during the operations phase is gratefully
acknowledged from the Istituto Nazionale di Astrofisica in Italy and
the Centre National d'\'Etudes Spatiales in France.

H.A. acknowledges supports provided by the NASA sponsored {\it Fermi}
Contract NAS5-00147 and by
Kavli Institute for Particle Astrophysics and Cosmology (KIPAC).
This work was supported by the research grant of the Chungbuk National University in 2016.

\bibliographystyle{apj}
\bibliography{GBINARY,BLLacs,FERMIBASE,PSRBINARY,STATISTICS}

\end{document}